\documentclass[aps,prd,superscriptaddress,nofootinbib]{revtex4-2}

\usepackage{supertabular} 
\usepackage{placeins}
\usepackage{epsfig}
\usepackage{graphicx}
\usepackage{tensor}

\usepackage{float}
\usepackage{mathtools}
\usepackage{array}
\usepackage{dcolumn}

\usepackage{latexsym}
\usepackage{amsmath}
\usepackage{amssymb}
\usepackage{amsfonts}
\usepackage{bm}
\usepackage{physics}

\usepackage{color}
\definecolor{purple}{rgb}{0.5,0,0.5}
\definecolor{blue}{rgb}{0.0,0,0.9}
\definecolor{prdblue}{rgb}{0.133,0.118,0.498}
\usepackage[colorlinks=true, pdfstartview=FitV, linkcolor=prdblue, citecolor= prdblue, urlcolor=prdblue]{hyperref}

\allowdisplaybreaks
\begin{document}


\title{Gravitational Instability of Cold Axion Dark Matter Beyond Minimal Coupling}

\author{Bryan Cordero-Patino}
\email[]{bcordero@unizar.es}
\affiliation{CAPA \& Departamento de Física Teórica, Universidad de Zaragoza, C. Pedro Cerbuna 12, 50009 Zaragoza, Spain}

\author{\'Alvaro Duenas-Vidal}
\email[]{alvaro.duenas@epn.edu.ec}
\affiliation{Departamento de Física, Escuela Politécnica Nacional, Quito 170143, Ecuador}

\author{Jorge Segovia}
\email[]{jsegovia@upo.es}
\affiliation{Departamento de Sistemas F\'isicos, Qu\'imicos y Naturales, Universidad Pablo de Olavide, Avenida Rectora Rosario Valpuesta 1, 41089, Dos Hermanas, Sevilla, Spain.}

\date{\today}

\begin{abstract}

Following recent observations of unexpectedly massive galaxies at high redshift by the James Webb Space Telescope, the revision of Dark Matter (DM) models—specifically regarding how DM couples to gravity—has become a central topic in Cosmology. In this paper, we develop an axion model for Cold Dark Matter (CDM) with a non-minimal coupling to gravity. From this framework, we evaluate the gravitational instability in the hydrodynamic limit. Our results demonstrate how gravitational instabilities depend on the non-minimal coupling parameter, providing a novel tool for aligning with the high-redshift observations from the James Webb Space Telescope.

\end{abstract}


\maketitle


\section{INTRODUCTION}
\label{sec:introduction}

As one of the fundamental components of our universe, Dark Matter (DM) is responsible for the early formation of cosmic structures. Consequently, it dictates the statistics of galaxies observed today, whose origin remains poorly understood. With an extremely weak (or even non-existent) coupling to radiation, it is the first species to have decoupled in the thermal history of the Universe and, consequently, the first to be susceptible to gravitational collapse. In this way, DM provided gravitational potential wells that ultimately served as scaffolding for ordinary matter to form the first galaxies, driven by a hierarchical bottom-up process. Providing any clue regarding the characteristics of the gravitational collapse of DM in the Early Universe is, therefore, a goal toward which Cosmology must direct a significant portion of its efforts. This is especially mandatory after recent observations from the James Webb Space Telescope, which reveal unexpected behavior for structure formation in the Early Universe~\cite{Labbe2023, BoylanKolchin2023, Finkelstein2023}.

The early gravitational collapse of DM depends heavily on the model used for its description: mass, self-interaction, temperature, and coupling to gravity. Regarding temperature, recent observations from the James Webb Space Telescope have ruled out Hot Dark Matter and Warm Dark Matter models, while significantly constraining Fuzzy Dark Matter models and underscoring the need to explore Cold Dark Matter models in greater detail. Thus, it becomes imperative to investigate gravitational collapse in Cold Dark Matter (CDM) models, considering couplings to gravity beyond minimal coupling. In this paper, we focus on the non-relativistic axion field as the underlying microscopic model for CDM. Among the proposed DM candidates, the axion continues to receive considerable attention in both experimental and theoretical investigations~\cite{Battat::2024, Jackson::2023, ParticleDataGroup::2022}.

To make this paper self-contained, let us briefly review the axion framework. The axion was originally introduced as a byproduct of the Peccei-Quinn mechanism, which continues to be the most relevant solution to the strong CP problem of Quantum Chromodynamics (QCD)~\cite{Peccei::1977, Peccei_2::1977}. Many of the axion's fundamental properties that make it a compelling DM candidate arise from its connection to the QCD sector. The small mass of the axion $m_a$ may appear, at first glance, to be in contradiction with its role as DM. Nevertheless, non-thermal production mechanisms, such as the vacuum misalignment mechanism, the kinetic misalignment mechanism, summit-landing branches, and the decay of axionic cosmic strings, can generate a sufficiently large axion population to account for a substantial fraction, or potentially all, of the observed DM density~\cite{ParticleDataGroup::2022, Irastorza::2021,Eroncel_2022,Chathirathas::2026}. The axion mass is related to the topological susceptibility of QCD $\chi$ and the decay constant of the axion $f_a$ through $m_a = \sqrt{\chi}/f_a$. The most precise computation to date yields~\cite{ParticleDataGroup::2022, snowmass_axions::2022}:

\begin{equation}
m_a = 5.691(51) \left(\frac{10^{12} \text{GeV}}{f_a} \right) \mu \text{eV} \,,
\end{equation}

\noindent
giving a viable DM mass window for the axion of $m_a \in (10^{-6} - 10^{-3}) \, \text{eV}$~\cite{Irastorza::2021}. At energy scales below which QCD non-perturbative effects become significant, the axion can be described by a real scalar field $\phi$ governed by the Klein--Gordon action,

\begin{equation}
S = - \int d^4 x \, \left( \frac{1}{2} \partial_\mu \phi \partial^\mu \phi + V(\phi) \right) \,.
\label{eq::axion-rltvs_action}
\end{equation}

\noindent
The potential $V(\phi)$ contains the axion self-interactions and can be derived either within the instaton gas approximation or, alternatively, using chiral perturbation theory~\cite{Luzio::2020}. Since the potential is even under $\phi\rightarrow-\phi$, it can be expanded in even powers of $\phi$ around 0,

\begin{equation}
V(\phi) = \frac{1}{2} m_a^2 \phi^2 + \left(m_a f_a \right)^2 \sum_{n=2}^{\infty} \frac{\lambda_{2n}}{(2n)!} \left(\frac{\phi}{f_a} \right)^{2n} \,,
\label{eq::axion-rltvs_potential}
\end{equation}

\noindent
where $\lambda_{2n}$ are dimensionless coupling constants of $\mathcal{O}(1)$. The dimensionless coupling constant for the 4-axion vertex $\lambda_4$ is particularly relevant, Since pairwise axion interactions provide the leading contribution to the self-interaction potential, we retain only the quartic term throughout this work. For convenience, we introduce the effective coupling constant

\begin{equation}
    \Bar{\lambda}_4 \equiv \lambda_4 \frac{m_a^2}{f_a^2} \, .
\end{equation}

\noindent
where $z = m_u/m_d$ is the up-to-down quark mass ratio. It is easy to see that $\lambda_4 < 0$ regardless of the value of $z$, i.e., the axion pair interactions are attractive. Based on next-to-leading order analysis in chiral effective field theory, $z = 0.48(3)$ is the accepted value for this ratio\cite{Braaten::2019}. Since the axion pair-interaction is the most relevant interaction, only this term will be maintained through this work. It is convenient to introduce the notation

\begin{equation}
    \Bar{\lambda}_4 \equiv \lambda_4 \frac{m_a^2}{f_a^2} \, .
\end{equation}

Tackling the problem of gravitational collapse in axion-based CDM is no simple task. Because CDM is non-relativistic, it is necessary to obtain a low-energy description of the axion field commonly known as a non-relativistic effective field theory (NREFT)~\cite{Namjoo::2017, Salehian::2020, Salehian::2021, Cordero::2023,Cordero-Patino:2026lxh}. Additionally, the evolution of axion structures may be influenced by the expansion of the Universe and, thus, it is necessary to incorporate General Relativity to the axion description. This can be solved naturally by means of minimal coupling  of the scalar axion field to spacetime curvature. The resulting Einstein-Klein-Gordon system provides the starting point for describing self-gravitating axion configurations and their formation. By applying cosmological perturbation theory with the Newtonian gauge and subsequently taking the non-relativistic limit, one obtains the Gross-Pitaevskii-Poisson equations ~\cite{Jackson::2023,Cordero-Patino:2026lxh}. Then, following the work of Ref.~\cite{Suarez::2018}, the Madelung transformation can be performed in the resulting system of equations to obtain a hydrodynamic representation of the system from which the dispersion relation and Jeans instability of an axion cloud can be calculated more freely.

Beyond the path we have just outlined to reach Jeans instability in axion-like CDM models, minimal coupling is not the most general interaction between a scalar field and gravity. From the perspective of effective field theory in curved spacetime, scalar fields can acquire couplings to curvature invariants through quantum corrections and renormalization, even when such interactions are not present in the classical action~\cite{Birrell:1982ix, Buchbinder:1992rb}. Non-minimal scalar-curvature interactions therefore provide a natural extension of the conventional Einstein-Klein-Gordon description. The existence of extra coupling terms beyond minimal coupling can be of crucial importance in the fate of the gravitational collapse~\cite{Chavanis::2011, Eby::2016}. Motivated by this observation, we investigate a non-minimal interaction of the form $R\phi^4$, with $R$ denoting the Ricci scalar. This term retains the scalar self-interaction structure while introducing an additional curvature-dependent contribution to the dynamics of the axion field. Because the Ricci scalar depends on the local gravitational environment, the resulting contribution can modify the effective interaction experienced in an axion condensate and may provide either attractive or repulsive corrections.

The structure of this manuscript is as follows. In Section~\ref{sec::SETUP} a summary of the results obtained in Ref.~\cite{Cordero-Patino:2026lxh} are described. In Section~\ref{sec::HYDR_LIMIt} the Madelung transformation is performed to the Gross-Pitaevskii-Poisson equations modified by the non-minimal coupling term to obtain a hydrodynamic representation. In Section~\ref{sec::DISPERTION_RLT}, the dispersion relation and Jeans length are computed for a static Universe. Finally, we summarize our results and outline possible directions for future work in Section~\ref{sec:summary}.


\section{SETUP FOR NON-RELATIVISTIC AND NON-MINIMAL COUPLED AXION FIELD}
\label{sec::SETUP}
The action describing axionic matter minimally coupled to gravity is

\begin{equation}
    S_m = - \int d^4 x \, \sqrt{-g} \left(\frac{1}{2} \, \partial^\mu \phi \partial_\mu \phi + V(\phi) \right) \,,
    \label{eq::action_m_1}
\end{equation}

\noindent
where $g = \det (\tensor{g}{_\mu_\nu})$. Minimal coupling describes curvature effects only at the most basic level. Non-minimal couplings appear naturally in various settings such as QFT in curved spacetime, Higgs inflation models, modified gravity theories, and the low-energy superstring effective field theories. These additional modifications must remain diffeomorphism invariant and preserve the internal symmetries of the theory. They are typically constructed from the Riemann tensor $\tensor{R}{_\mu_\nu_\rho_\sigma}$ and its contractions, ensuring that they vanish in Minkowski spacetime.

Since the axion pair self-interactions are described by the $\phi^4$ term in $V(\phi)$, one can naturally extend the theory by adding a curvature-dependent contribution proportional to
$R \phi^4$, where $R$ is the curvature scalar. Unlike the standard $R \phi^2$ coupling, which mainly shifts the effective mass, the $R \phi^4$ operator directly modifies the quartic interaction relevant for condensate dynamics. As a result, the total action becomes

\begin{align}
S_{tot} =& - \int \, d^4 x \, \sqrt{-g} \left(\frac{1}{2} \, \partial^{\mu} \phi \, \partial_{\mu} \phi + V (\phi) + \xi R \phi^4 \right) +\frac{1}{16\pi G} \int d^4x \sqrt{-g} \, R \,,
\label{eq::action_non_min_coupling}
\end{align}
where $\xi$ is a constant.

It is compelling to study the evolution of axionic matter density perturbations in a cosmological background. For this end, cosmological perturbation theory is employed using the Newtonian gauge metric. A particularly relevant study case is the evolution of spherically symmetric matter perturbations. In this context, it is convenient to express the spatial components in spherical coordinates; thereby, the Newtonian gauge line element takes the form
\begin{equation}
ds^2 = a^2(\eta) \left[-\left(1+2\,\Phi\right) d\eta^2 + \left(1-2\Psi\right) \left(dr^2 + r^2 d\theta^2 + r^2 \sin^2(\theta) d\varphi^2 \right) \right] \,,
\label{eq::Newtonian_gauge_spherical}
\end{equation}

\noindent
where $\eta$ is conformal time, $\Phi(x)$ and $\Psi(x)$ are scalars which follow that $|\Phi|,|\Psi| \ll 1$, and $a(\eta)$ is the dimensionless scale factor. 

Before decomposing the scalar field as $\phi(x) = \overline{\phi}(\eta) + \delta \phi(x)$ to derive the equations governing the homogeneous background $\overline{\phi}$ and the perturbation $\delta \phi$, it is useful to first obtain the low-energy limit of the theory, since DM is predominantly non-relativistic. This can be done by mapping the real field $\phi$ to a complex field $\psi$ through an exact transformation,

\begin{align}
\phi(x) &= \frac{1}{\sqrt{2m_a}} \left( e^{-im_at} \, \psi(x) + e^{im_at} \, \psi^*(x) \right) \,, \\
\partial_t \phi(x) &= -i\sqrt{\frac{m_a}{2}} \left( e^{-im_at} \, \psi(x) - e^{im_at} \, \psi^*(x) \right) \,. \label{eq::transf_Salehian}
\end{align}

The equations governing the evolution of the system are a modified version of the Gross-Pitaevskii-Einstein equations. For further details on their computation, see Ref.~\cite{Cordero-Patino:2026lxh}. It is notable that if the axion field remains spherically symmetric throughout its evolution, the equations of motion naturally enforce $\Psi = \Phi$, thus mirroring an absence of anisotropic stress. This assumption will be adopted throughout the remainder of this work. Under these conditions, the Euler-Lagrange equation simplifies to

\begin{align}
    i \partial_\eta \psi =& \; a {\left(m_a \Phi - \frac{3i}{2} H \right)} \psi + 2i \, \partial_\eta \Phi \, \psi - \frac{\nabla^2 \psi}{2am_a} + \left( \frac{a \Bar{\lambda}_4}{8 m_a^2} \left(1 + 2\, \Phi \right) + \frac{9\xi}{a m_a^2} (a^2 H^2 - \Upsilon) \right) \psi |\psi|^2 \nonumber \\ 
    &- \frac{6\xi}{a m_a^2} \left( 12 a H \, \partial_\eta \Phi + 3 \, \partial_\eta^2 \Phi - \nabla^2 \Phi \right) \psi |\psi|^2 \,.
    \label{eq::GPPE_non_minimal}
\end{align}

\noindent
where $H(\eta) = \frac{1}{a^2} \frac{da}{d\eta}$ is the Hubble parameter and $\Upsilon(\eta) = a^2 H^2 - \frac{2}{a} \frac{d^2 a}{d \eta^2}$. Additionally, the Einstein field equations have to be considered. In this work the only relevant equations are the double temporal equation,

\begin{align}
    \left(\frac{1}{8\pi G} - \frac{3\xi}{m_a^2} |\psi|^4 \right) \left(- 3 \, H^2 \left(1 - 2\, \Phi \right) + \frac{6H}{a} \partial_\eta \Phi - \frac{2}{a^2} \, \nabla^2 \Phi \right)  + m_a |\psi|^2  + \frac{\Bar{\lambda}_4}{16m_a^2} |\psi|^4 -m_a |\psi|^2 \, \Phi - \frac{3 \xi}{a^2m_a^2} \nabla^2 |\psi|^4 = 0 \,. \label{eq::Einst-eta_eta}
\end{align}

\noindent
and the trace of the spatial equations,

\begin{align}
    &\left(\frac{1}{8\pi G} - \frac{3\xi}{m_a^2} |\psi|^4 \right) \left(\frac{3}{a^2} \Upsilon \left(1 - 2\, \Phi \right) + \frac{6}{a^2} \, \partial_\eta^2 \Phi + \frac{18H}{a} \partial_\eta \Phi \right) + \frac{3 \Bar{\lambda}_4}{16m_a^2} |\psi|^4 \nonumber \\
    &+3 m_a |\psi|^2 \, \Phi - 72 \xi \Phi |\psi|^4  +\frac{6 \xi}{a^2 m_a^2} {\left[ 3 \, |\psi|^2 \left(\psi \nabla^2 \psi^* + \psi^* \nabla^2 \psi \right) - \nabla^2 |\psi|^4 \right]} = 0 \,. \label{eq::Einst-trace_spatial}
\end{align}


\section{HYDRODYNAMIC LIMIT}
\label{sec::HYDR_LIMIt}

When studying axionic structures, it is customary to reformulate the non-relativistic theory in terms of hydrodynamic variables. This approach reshapes the field equations into a set of fluid-like equations whose physical interpretation is more intuitive. The hydrodynamic description is obtained through the Madelung transformation,

\begin{equation}
    \psi(x) = \sqrt{\rho(x)} \; e^{iS(x)} \,,
    \label{eq::Madelung_transf}
\end{equation}

\noindent
where $\rho(x) = |\psi(x)|^2$ is the axion number density and $S(x)$ is the action from which the velocity field, 

\begin{equation}
    \Vec{v}(x) = \frac{\nabla S(x)}{a m_a}\,,
\end{equation}

\noindent
is defined. It is important to note that the velocity field is irrotational, i.e., $\nabla \times \Vec{v} = 0$.

Applying Eq.~(\ref{eq::Madelung_transf}) to Eq.~(\ref{eq::GPPE_non_minimal}) yields two equations: one for the imaginary part,

\begin{align}
    \partial_\eta \rho + \nabla \cdot \left(\rho \Vec{v} \right) =& - 3a H \rho + 4 \rho \, \partial_\eta \Phi \,, \label{eq::continuity}
\end{align}

\noindent
which has the form of a continuity equation; and one for the real part,

\begin{align}
    0  =& \; \partial_\eta S + \frac{(\nabla S)^2}{2am_a} + a m_a \Phi -\frac{1}{4am_a} \left( \frac{\nabla^2 \rho}{\rho} - \frac{1}{2} \frac{(\nabla \rho)^2}{\rho^{2}} \right) + \left( \frac{a \Bar{\lambda}_4}{8 m_a^2} \left(1 + 2\, \Phi \right) + \frac{9\xi}{a m_a^2} (a^2 H^2 - \Upsilon) \right) \rho \nonumber \\
    & - \frac{6\xi}{a m_a^2} \left( 12 a H \, \partial_\eta \Phi + 3 \, \partial_\eta^2 \Phi- \nabla^2 \Phi \right) \rho \,. \label{eq::Hamilton-Jacobi}
\end{align}

\noindent
which is analogous to the Hamilton-Jacobi equation for an irrotational fluid. The fourth term of Eq.~(\ref{eq::Hamilton-Jacobi}) is homologous to the Bohm quantum potential that is characteristic feature of the hydrodynamic representation. It is often expressed as

\begin{equation}
    Q = -\frac{1}{2m_a} \frac{\nabla \sqrt{\rho}}{\sqrt{\rho}} =-\frac{1}{4m_a} \left( \frac{\nabla^2 \rho}{\rho} - \frac{1}{2} \frac{(\nabla \rho)^2}{\rho^{2}} \right) \,.
\end{equation}

\noindent
Likewise, Eq.~(\ref{eq::Madelung_transf}) has to be implemented on the Einstein field equations. The double temporal equation becomes

\begin{align}
    \left(\frac{1}{8\pi G} - \frac{3\xi}{m_a^2} \rho^2 \right) \left(- 3 \, H^2 \left(1 - 2\, \Phi \right) + \frac{6H}{a} \partial_\eta \Phi - \frac{2}{a^2} \, \nabla^2 \Phi \right)  + m_a \rho  + \frac{\Bar{\lambda}_4}{16m_a^2} \rho^2 -m_a \rho \, \Phi - \frac{3 \xi}{a^2m_a^2} \nabla^2 \rho^2 = 0 \,.
\end{align}

\noindent
The trace of the spatial equations results in

\begin{align}
    &\left(\frac{1}{8\pi G} - \frac{3\xi}{m_a^2} \rho^2 \right) \left(\frac{3}{a^2} \Upsilon \left(1 - 2\, \Phi \right) + \frac{6}{a^2} \, \partial_\eta^2 \Phi + \frac{18H}{a} \partial_\eta \Phi \right) + \frac{3 \Bar{\lambda}_4}{16m_a^2} \rho^2 \nonumber \\
    &+3 m_a \rho \, \Phi - 72 \xi \Phi \rho^2  +\frac{6 \xi}{a^2 m_a^2} {\left( 3 \, \rho \, \nabla^2 \rho - \nabla^2 \rho^2 \right)} = 0 \,.
\end{align}

The next step is to differentiate the background evolution from the perturbations. Throughout this work, an overbar denotes homogeneous background quantities, while the prefix $\delta$ represents the first-order perturbations. Setting $\rho(x) = \Bar{\rho}(\eta)$, $S(x) = \Bar{S}(\eta)$, and $\Phi(x) = 0$, the continuity and Hamilton-Jacobi equations become

\begin{align}
    \partial_\eta \Bar{\rho} + 3a H \, \Bar{\rho} =& \; 0 \,, \\
    \partial_\eta \Bar{S} + \left( \frac{a \Bar{\lambda}_4}{8 m_a^2} + \frac{9\xi}{a m_a^2} (a^2 H^2 - \Upsilon) \right) \Bar{\rho} =& \; 0 \,.
\end{align}

\noindent
Furthermore, the Einstein equations become

\begin{align}
    \left(\frac{1}{8\pi G} - \frac{3\xi}{m_a^2} \Bar{\rho}^2 \right) \left(- 3 \, H^2 \right) + m_a \Bar{\rho}  + \frac{\Bar{\lambda}_4}{16m_a^2} \Bar{\rho}^2 =& \; 0 \,, \\
    \left(\frac{1}{8\pi G} - \frac{3\xi}{m_a^2} \Bar{\rho}^2 \right) \left(\frac{\Upsilon}{a^2} \right) + \frac{ \Bar{\lambda}_4}{16m_a^2} \Bar{\rho}^2  =& \; 0 \,.
\end{align}

Afterwards, its necessary to set $\rho(x) = \rho(\eta) + \delta \rho(x)$ and $S(x) = S(\eta) + \delta S(x)$ with $\delta \rho \ll \rho$ and $|\delta S| \ll |S|$. After subtracting the background evolution, the continuity and Hamilton-Jacobi equations become 

\begin{align}
    \partial_\eta \left(\delta \rho\right) + \frac{\Bar{\rho}}{am_a} \nabla^2 \left(\delta S\right) =& - 3a H \delta \rho + 4 \Bar{\rho} \, \partial_\eta \Phi  \,,
\end{align}

\begin{align}
    0  =& \; \partial_\eta \left(\delta S\right) + a m_a \Phi -\frac{1}{4am_a} \frac{\nabla^2 \delta \rho}{\Bar{\rho}} + \frac{a \Bar{\lambda}_4}{4 m_a^2} \Bar{\rho} \Phi + \left( \frac{a \Bar{\lambda}_4}{8 m_a^2} + \frac{9\xi}{a m_a^2} (a^2 H^2 - \Upsilon) \right) \delta \rho \nonumber \\
    & - \frac{6\xi \Bar{\rho}}{a m_a^2} \left( 12 a H \, \partial_\eta \Phi + 3 \, \partial_\eta^2 \Phi - \nabla^2 \Phi \right)  \,.
    \label{eq::Hamilton_Jacobi_linear_pertb}
\end{align}

Additionally, the Einstein equations become

\begin{align}
    &\frac{1}{4\pi \mathcal{G}} \left(3 \, a^2 H^2 \Phi + 3aH \partial_\eta \Phi - \nabla^2 \Phi \right) -a^2 m_a \Bar{\rho} \, \Phi + a^2 \left[m_a + \frac{\Bar{\rho}}{m_a^2} \left( \frac{\Bar{\lambda}_4}{8} + \frac{6 \xi}{a^2} \left(3 a^2 H^2 - \nabla^2\right) \right) \right] \delta \rho = 0 \,.
\end{align}

\begin{align}
    &\frac{1}{8\pi \mathcal{G}} \left(\partial_\eta^2 \Phi + 3 aH \, \partial_\eta \Phi - \Upsilon \, \Phi \right) + \frac{1}{2} a^2 m_a \Bar{\rho} \, \Phi - 12 \xi a^2 \Bar{\rho}^2 \Phi + \frac{a^2 \Bar{\rho}}{m_a^2} \left( \frac{\Bar{\lambda}_4}{16} - \frac{\xi}{a^2} \left( 3 \Upsilon - \nabla^2 \right) \right) \delta \rho= 0 \,,
\end{align}

\noindent
where

\begin{align}
    8\pi \, \mathcal{G}(\eta) =& \; {\left( \frac{1}{8\pi G} - \frac{3 \xi}{m_a^2} \Bar{\rho}^2 \right)}^{-1} .
\end{align} 


\section{DISPERSION RELATION AND JEANS LENGTH IN A STATIC UNIVERSE}
\label{sec::DISPERTION_RLT}

If the fragmentation timescale of the axion cloud is much shorter than the Hubble timescale, the expansion of the Universe can be safely neglected. In this limit, the background spacetime is effectively static, so that $a = 1$, $H=0$, and $\Upsilon = 0$. It is then convenient to use the following adimensional variables:

\begin{align}
    \mu \equiv \frac{\delta \rho}{\Bar{\rho}} \,, && \sigma \equiv \frac{\partial_\eta (\delta S)}{m_a}\,.
\end{align}

\noindent
Hence, the equations of motion of the perturbation become

\begin{align}
    \partial_\eta \mu + \nabla \cdot \Vec{v} =& \; 4 \, \partial_\eta \Phi  \,,
    \label{eq::continuity_static_universe}
\end{align}

\begin{align}
    m_a \, \sigma + \left(m_a  + \frac{\Bar{\lambda}_4}{4 m_a^2} \Bar{\rho} \right)  \Phi -\frac{1}{4m_a} \nabla^2 \mu + \frac{ \Bar{\lambda}_4}{8 m_a^2} \Bar{\rho} \, \mu - \frac{6\xi \Bar{\rho}}{m_a^2} \left( 3 \, \partial_\eta^2 \Phi - \nabla^2 \Phi \right) &= 0  \,,
    \label{eq::Hamilton_Jacobi_static_universe}
\end{align}

\begin{align}
    &- \frac{1}{4\pi \mathcal{G}} \nabla^2 \Phi - m_a \Bar{\rho} \, \Phi + \Bar{\rho} \left[m_a + \frac{\Bar{\rho}}{m_a^2} \left( \frac{\Bar{\lambda}_4}{8} - 6 \xi \, \nabla^2 \right) \right] \mu = 0 \,,
    \label{eq::double_temporal_static_universe}
\end{align}

\begin{align}
    &\frac{1}{8\pi \mathcal{G}} \partial_\eta^2 \Phi + \frac{1}{2} m_a \Bar{\rho} \, \Phi - 12 \xi \Bar{\rho}^2 \Phi + \frac{ \Bar{\rho}^2}{m_a^2} \left( \frac{\Bar{\lambda}_4}{16} + \xi \, \nabla^2 \right) \mu = 0 \,.
    \label{eq::trace_spatial_static_universe}
\end{align}

\noindent
Combining Eqs.(\ref{eq::double_temporal_static_universe}) and~(\ref{eq::trace_spatial_static_universe}) is possible to obtain

\begin{align}
    \frac{1}{4\pi \mathcal{G}} \left( 3 \, \partial_\eta^2 \Phi - \nabla^2 \Phi \right) = - 2 m_a \Bar{\rho} \, \Phi + 72 \xi \Bar{\rho}^2 \Phi - \left(m_a \Bar{\rho} + \frac{\Bar{\lambda}_4 \Bar{\rho}^2}{2 m_a^2} \right) \delta\,.
    \label{eq::Hamilton_Jacobi_static_universe_2}
\end{align}

\noindent
Replacing on Eq.(\ref{eq::Hamilton_Jacobi_static_universe}),

\begin{align}
    0  =& \; m_a \, \sigma + \left(m_a  + \frac{\Bar{\lambda}_4}{4 m_a^2} \Bar{\rho} + \frac{48 \pi \mathcal{G} \xi \Bar{\rho}^2}{m_a^2} \left(m_a - 36 \xi \Bar{\rho} \right) \right)  \Phi -\frac{1}{4m_a} \nabla^2 \mu +  \, \left(\frac{ \Bar{\lambda}_4 \Bar{\rho}}{8 m_a^2} + \frac{24 \pi \mathcal{G} \xi \Bar{\rho}^2}{m_a} + \frac{12 \pi \mathcal{G} \xi \Bar{\lambda}_4 \Bar{\rho}^3}{m_a^4} \right) \mu \,.
\end{align}

By expanding $\mu =\mu_k \exp \left(i (\Vec{k} \cdot \Vec{x} - \omega \eta) \right) $, $\sigma =\sigma_k \exp \left(i (\Vec{k} \cdot \Vec{x} - \omega \eta) \right) $, and $\Phi =\Phi_k \exp \left(i (\Vec{k} \cdot \Vec{x} - \omega \eta) \right) $ into Eqs. (\ref{eq::continuity_static_universe}), (\ref{eq::double_temporal_static_universe}), and (\ref{eq::Hamilton_Jacobi_static_universe_2}), it is possible to show that the dispersion relation is 

\begin{align}
    \omega^2 =&\; \frac{ \left( \frac{1}{16\pi G m_a} - \frac{3 \xi \Bar{\rho}^2}{2m_a^3} \right) k^6 + a(\xi,\Bar{\rho}) k^4 - b(\xi,\Bar{\rho}) k^2 }{ \left( \frac{m_a}{4\pi G} + \frac{18 \xi \Bar{\rho}^2}{m_a} \right) k^2 +3 m_a^2 \Bar{\rho} + \frac{\Bar{\lambda}_4 \Bar{\rho}^2}{2 m_a}} \,,
    \label{eq::dispertion_relation}
\end{align}

\noindent
where

\begin{align}
    a(\xi,\Bar{\rho}) =& \; \frac{\Bar{\lambda}_4 \Bar{\rho}}{32 \pi G m_a^2} + \frac{3 \Bar{\lambda}_4 \xi  \Bar{\rho}^3}{4m_a^4} - \frac{\Bar{\rho}}{4} -\frac{288 \pi G \xi^2 \Bar{\rho}^4}{m_a^2} \frac{m_a - 36 \xi \Bar{\rho}}{m_a^2 - 24 \pi G \xi \Bar{\rho}^2} \,,\\
    b(\xi,\Bar{\rho}) =&\; m_a^2 \Bar{\rho}  + \frac{\Bar{\lambda}_4 \Bar{\rho}^2}{2 m_a} + \frac{\Bar{\lambda}_4^2 \Bar{\rho}^3}{32 m_a^4} + \frac{9 \pi G \xi \Bar{\rho}^3}{m_a^2 - 24 \pi G \xi \Bar{\rho}^2} \left[ \left(8m_a + \frac{\Bar{\lambda}_4 \Bar{\rho}}{m_a^2} \right) \left( m_a - 24 \xi \Bar{\rho} \right) + \frac{\Bar{\lambda}_4 \Bar{\rho}}{m_a} \right] \,.
\end{align}

\noindent
It is convenient to define 

\begin{equation}
    \xi_\text{limit} = \frac{m_a^2}{24 \pi G \Bar{\rho}^2} \,.
\end{equation}

\begin{figure}
    \centering
    \includegraphics[width=0.75\linewidth]{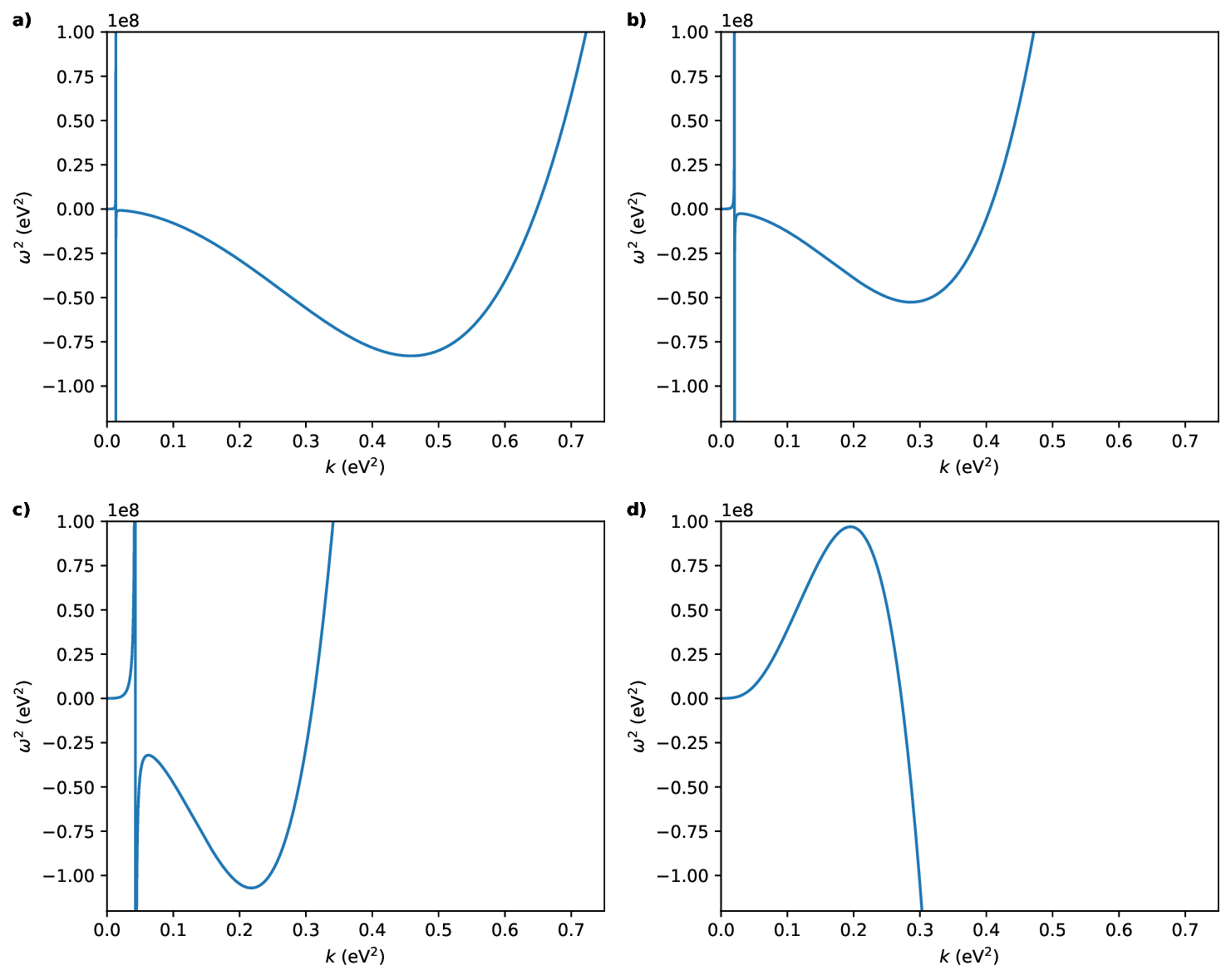}
    \caption{Comparison of the dispersion relation for different values of $\xi$. a) $\xi = \xi_\text{limit}(1 - 10^{-0.25})$, b) $\xi = 0$, c) $\xi = \xi_\text{limit}(1 - 10^{0.1})$, and d) $\xi = \xi_\text{limit}(1 - 10^{0.14})$.}
    \label{fig::dispersion_different_omega}
\end{figure}

In Fig.~\ref{fig::dispersion_different_omega} it is shown a comparison of the dispersion relation for different values of $\xi$. It was used $f_a = 10^{21}$ eV, $z= 0.48$, and $\Bar{\rho} = 2.5 \times 10^{-3} \frac{m_a}{4 \pi G}$. Due to the denominator in Eq.(\ref{eq::dispertion_relation}), there appears a discontinuity in all curves. When $\xi$ is negative, the shape of the curve is maintained. However, as $\xi$ becomes mores negative, the dispersion relation's depth widens and the point where the curve intersects with the $x$-axis shifts to the left. Meanwhile, when $\xi$ is positive the shape of the curve is mirrored.

\begin{figure}
    \centering
    \includegraphics[width=0.5\linewidth]{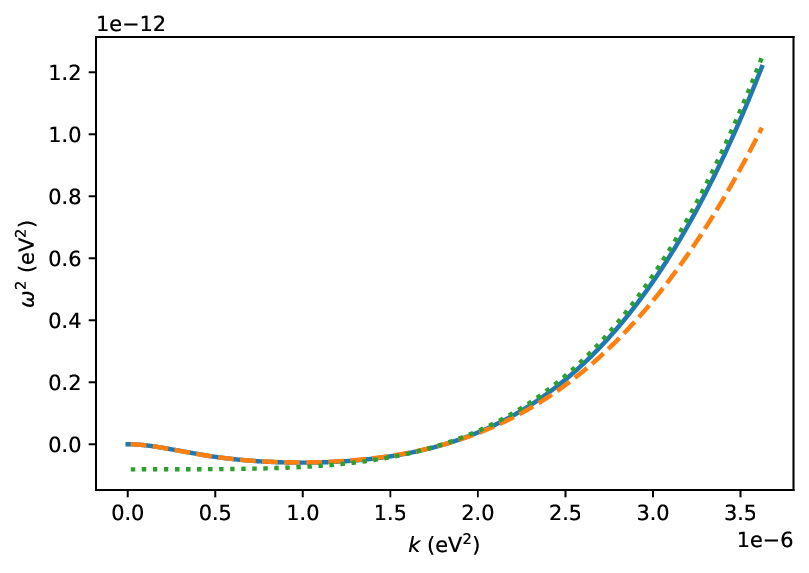}
    \caption{Comparison of the dispersion relation with the results of Ref.\cite{Suarez::2018}. In blue is the expression obtained in this work. In orange is the relativistic expression obtained in Ref.\cite{Suarez::2018}. In green is the non-relativistic expression obtained in Ref.\cite{Suarez::2018}.}
    \label{fig::dispersion_different_papers} 
\end{figure}

In Fig.~\ref{fig::dispersion_different_papers} it is shown a comparison of the dispersion relation obtained in this paper with the expressions obtained in Ref.\cite{Suarez::2018}. In this graph it is used $f_a = 10^{21}$ eV, $\lambda_4= 0$, $\xi=0$, and $\Bar{\rho} = 2.5 \times 10^{-3} \frac{m_a}{4 \pi G}$. It is clear that the expression obtained in this work closely follows the relativistic result obtained in Ref.\cite{Suarez::2018}. This is somewhat surprising because Ref.\cite{Suarez::2018} uses a complex field as the axion field instead of a real field. The discontinuity that appears on Figure \ref{fig::dispersion_different_omega} is not present here because the values used are such that the denominator of Eq.(\ref{eq::dispertion_relation}) does not crosses 0 at any point.

\begin{figure}
    \centering
    \includegraphics[width=0.75\linewidth]{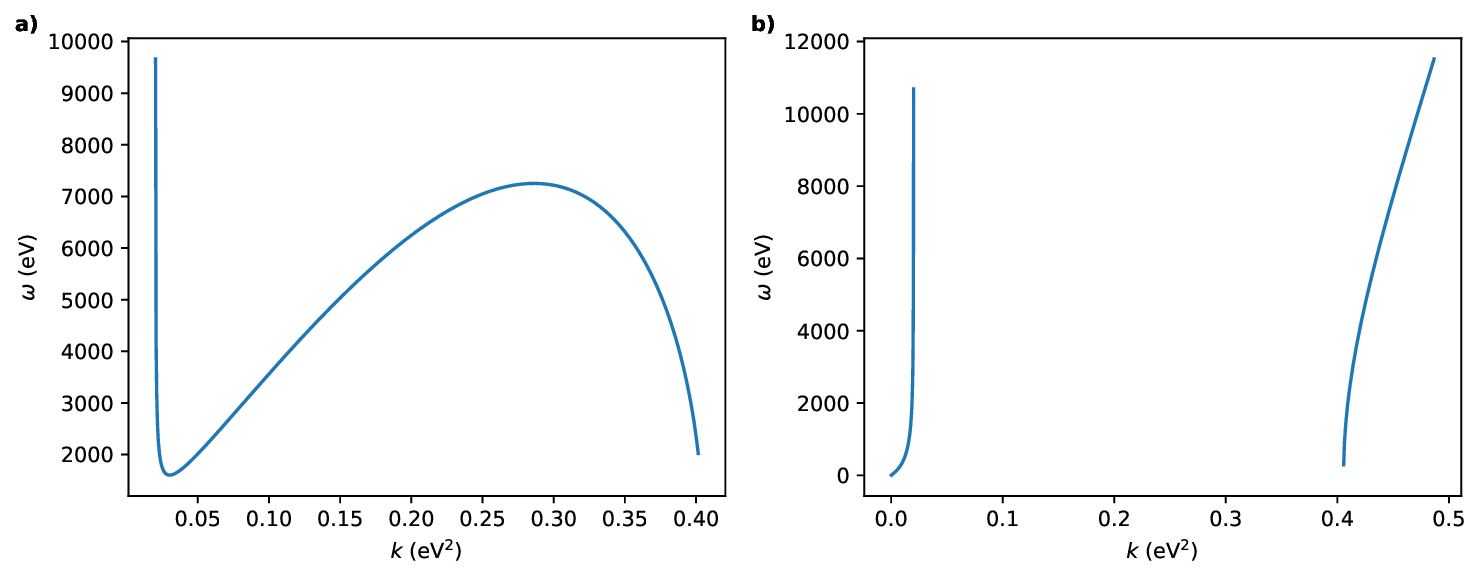}
    \caption{Absolute value of the branchs of $\omega$. a) Imaginary branch b) Real branch.}
    \label{fig::branches_imaginary}
\end{figure}

In Fig.~\ref{fig::branches_imaginary}, the imaginary and real branches of $\omega$ are shown, using $f_a = 10^{21}$ eV, $z= 0.48$, $\xi=0$, and $\Bar{\rho} = 2.5 \times 10^{-3} \frac{m_a}{4 \pi G}$.

Taking $\omega = 0$ in Eq.(\ref{eq::dispertion_relation}) its possible to calculate the Jeans length

\begin{align}
    k_J^2 =& \left( \frac{1}{8\pi G} - \frac{3 \Bar{\rho}^2 \xi}{m_a^2} \right)^{-1} \left(c(\xi, \Bar{\rho}) \pm \sqrt{c^2(\xi, \Bar{\rho}) + d(\xi, \Bar{\rho})} \right) \,,
    \label{eq::Jeans_length}
\end{align}

\noindent
where
\begin{align}
    c(\xi, \Bar{\rho}) =& \; \frac{m_a \Bar{\rho}}{4} - \frac{ \Bar{\lambda}_4 \Bar{\rho}}{32\pi G m_a} - \frac{3 \Bar{\lambda}_4 \Bar{\rho}^3}{4 m_a^3} \xi + {\left( \frac{1}{8\pi G} - \frac{3 \Bar{\rho}^2 \xi}{m_a^2} \right)}^{-1} \left( \frac{36 \Bar{\rho}^4}{m_a^2} \xi^2 -  \frac{1296 \Bar{\rho}^5}{m_a^3} \xi^3 \right) \,, \\
    d(\xi, \Bar{\rho}) =& - \left( 432 \Bar{\rho}^4 + \frac{54 \Bar{\lambda}_4 \Bar{\rho}^5}{m_a^3} \right) \xi^2 + \left( 12 m_a \Bar{\rho}^3 + \frac{3 \Bar{\lambda}_4 \Bar{\rho}^4}{2m_a^2} - \frac{3\Bar{\lambda}_4^2 \Bar{\rho}^5}{16 m_a^5} \right) \xi \nonumber \\
    &+ \frac{1}{8\pi G} \left( 2m_a^3 \Bar{\rho} + \Bar{\lambda}_4 \Bar{\rho}^2 + \frac{\Bar{\lambda}_4^2 \Bar{\rho}^3}{16 m_a^3} \right) \,.
\end{align}

\begin{figure}
    \centering
    \includegraphics[width=0.75\linewidth]{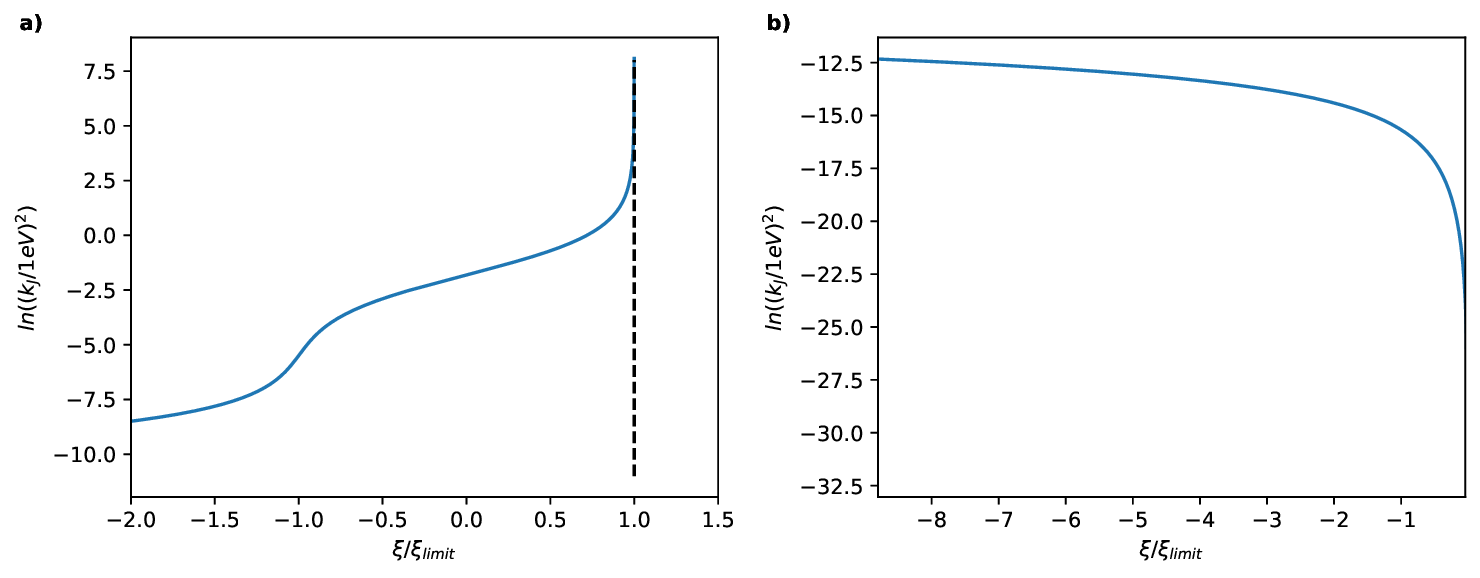}
    \caption{Jeans length for different autointeraction cases (on blue). The dashed blacked line represents the limit above which no Jeans length exists. The y-axis in both cases is in logarithmic scale. a)  $z=0.48$ and b) $\lambda_4 =0$.}
    \label{fig::Jeans_autointer}
\end{figure}

Graphs in Fig.\ref{fig::Jeans_autointer}  show how the Jeans length is modified as a function of $\xi$. Both graphs use $f_a = 10^{21}$ eV and $\Bar{\rho} = 2.5 \times 10^{-3} \frac{m_a}{4 \pi G}$. Left graph uses $z=0.48$, while the right graph uses $\lambda_4 = 0$. When self-interaction is present, $k_J$ decreases with decreasing $\xi$. Conversely, in the absence of self-interaction, $k_J$ increases as $\xi$ decreases. This shows that both $\lambda_4$ and $\xi$ are important when obtaining the limit where axion structures begin to form. Additionally, in both cases beyond $\xi_\text{limit}$, no physical value of $k_J$ exists; hence, this value of $\xi$ was designated as a limit.

\section{CONCLUSIONS}
\label{sec:summary}

In this paper, we have developed a model for axion CDM with a $\xi R \Phi^4$ non-minimal coupling to gravity, where $\xi$ is a real parameter. In the hydrodynamic limit, we demonstrate that the Jeans gravitational instability can be modulated by tuning both the non-minimal coupling $\xi$ and the self-interaction parameter $\lambda_4$ of the axion potential. Our results reveal a broad region in the parameter space that could be used to align with recent James Webb Space Telescope observations regarding the formation of unexpectedly massive galaxies in the Early Universe. In particular, the Jeans wavenumber in Eq.~\eqref{eq::Jeans_length} can be used to set a minimal mass threshold for early cosmic structures, assuming a hierarchical bottom-up process.

Note that we have proposed a framework in which the characteristic collapse timescale is assumed to be sufficiently small compared to the expansion rate of the Universe, allowing us to fix the scale factor $a = 1$. Future work must explore this assumption in greater detail by analyzing the collapse timescale within an expanding, radiation-dominated Universe. Further considerations arise from the relation between the axion decay constant $f_a$ and the energy scale of inflation, which determines whether the Peccei-Quinn symmetry is broken before or after inflation. These two scenarios have traditionally been associated with distinct phenomenological predictions, including differences in the formation and properties of axion miniclusters \cite{Eroncel_2023,Arvanitaki::2020,Gorghetto::2024}. However, recent studies have demonstrated that a substantial population of axion miniclusters can also emerge in the pre-inflationary scenario through kinetic misalignment and the so-called summit-landing branches \cite{Eroncel_2022,Chathirathas::2026}. These findings open up new avenues of investigation, particularly concerning how the non-minimal coupling considered in this work may influence the axion population and the axion structure formation dynamics.

Besides the early formation of cosmic structures, the collapse of CDM —which may involve a sub-structure fragmentation driven by post-inflationary spatial fluctuations in the axion field— can also manifest through the formation of compact \emph{dark} objects. This sub-fragmentation can subsequently give rise to localized, gravitationally bound configurations such as axion miniclusters and axion stars \cite{Braaten::2019, Chang::2024, Zhang::2018, Jackson_Kimball::2018}. The model developed here serves as a natural starting point to investigate the formation of such compact objects via gravitational collapse and subsequent fragmentation.


\begin{acknowledgments}
B.C-P. acknowledges financial support from PREP2024-002944, endorsed by MICIU/AEI/10.13039/501100011033 and by the FSE+. J.S.   acknowledges financial support from 
Ministerio Espa\~nol de Ciencia e Innovaci\'on under grant No. PID2022-140440NB-C22;
Junta de Andaluc\'ia under contract No. FQM-370 as well as PCI+D+i under the title: ``Tecnolog\'\i as avanzadas para la exploraci\'on del universo y sus componentes" (Code AST22-0001). A. D-V. acknowledges the financial support from project PIM-23-01 of the National Polytechnic School. 
\end{acknowledgments}


\bibliography{references}

\appendix

\end{document}